**Demystifying estimands in cluster-randomised trials**

Brennan C Kahan[1], Bryan S Blette[2], Michael O Harhay[1,3], Scott D Halpern[3], Vipul Jairath[4,5], Andrew Copas[1]*, Fan Li[6,7]*

*Contributed equally

[1] MRC Clinical Trials Unit at UCL, London, UK

[2] Department of Biostatistics, Vanderbilt University Medical Center, Nashville, USA

[3] Department of Biostatistics, Epidemiology, and Informatics, Perelman School of Medicine, University of Pennsylvania, Philadelphia, USA

[4] Department of Medicine, Division of Gastroenterology, Schulich school of Medicine, Western University.

[5] Department of Epidemiology and Biostatistics, Western University, London, ON, Canada

[6] Department of Biostatistics, Yale University School of Public Health

[7] Center for Methods in Implementation and Prevention Science, Yale University School of Public Health

**Correspondence to:** Brennan Kahan, b.kahan@ucl.ac.uk

**Abstract**

Estimands can help clarify the interpretation of treatment effects and ensure that estimators are aligned to the study's objectives. Cluster-randomised trials require additional attributes to be defined within the estimand compared to individually randomised trials, including whether treatment effects are *marginal* or *cluster-specific*, and whether they are *participant-* or *cluster-average*. In this paper, we provide formal definitions of estimands encompassing both these attributes using potential outcomes notation and describe differences between them. We then provide an overview of estimators for each estimand, describe their assumptions, and show consistency (i.e. asymptotically unbiased estimation) for a series of analyses based on cluster-level summaries. Then, through a re-analysis of a published cluster-randomised trial, we demonstrate that the choice of both estimand and estimator can affect interpretation. For instance, the estimated odds ratio ranged from 1.38 (p=0.17) to 1.83 (p=0.03) depending on the target estimand, and for some estimands, the choice of estimator affected the conclusions by leading to smaller treatment effect estimates. We conclude that careful specification of the estimand, along with an appropriate choice of estimator, are essential to ensuring that cluster-randomised trials address the right question.

# 1. Introduction

An estimand is a precise definition of the treatment effect investigators want to estimate [1-3]. Defining the estimand at the study outset helps to clarify the appropriate interpretation of treatment effects and ensure that statistical methods are aligned to the study's objectives (i.e. that statistical methods are chosen to estimate the *right* treatment effect) [1-11]. Because of the clarity that estimands provide, they are becoming increasingly popular in randomised trials. The standard framework for defining an estimand requires specification of the following five attributes: (i) the population of participants; (ii) the treatment conditions; (iii) the endpoint; (iv) the summary measure (e.g. odds ratio, difference, etc); and (v) how intercurrent events, such as treatment non-adherence, are handled. Importantly, the above attributes are defined in relation to the target treatment effect (e.g. the population to whom the treatment effect applies).

However, cluster-randomised trials (CRTs) (where groups of participants, such as schools or hospitals, are randomised instead of individual participants [12-18]) require the specification of additional attributes compared to individually randomised designs [4, 19]. For example, investigators must decide between *marginal* (sometimes called *population-averaged*) and *cluster-specific* (sometimes called *conditional*) treatment effects, which differ in whether outcomes are summarised overall or by cluster [14, 15, 20-22]. Separately, they must also decide between *participant-average* and *cluster-average* treatment effects, which differ in how participants are weighted [4].

Proper consideration of these attributes is important to ensure that CRTs are designed to answer the most clinically relevant question, as different estimands provide fundamentally different interpretations, and choosing the wrong estimand and/or estimator could provide misleading evidence. For instance, if interest lies in an intervention's effect across the population of participants (e.g. the number of participants that would be saved by using the intervention vs. using control), this is provided through a *participant-average* estimand. Hence, an estimator that targets a *cluster-average* estimand (such as an analysis of unweighted cluster-level summaries, which is a commonly recommended estimator in CRTs [14, 15, 18, 23, 24]), may provide a biased answer. It is therefore essential in CRTs to clearly define the estimand and then choose an appropriate estimator that targets this estimand.

The concept of marginal vs. cluster-specific treatment effects has been discussed previously [14, 15, 20-22], as has the issue around how to weight patients (i.e. participant- vs. cluster-average effects) [4, 19, 25-29]. However, to our knowledge, cluster-specific effects have not been formally defined using potential outcomes notation. Further, to our knowledge, these two issues have not been considered together, meaning there are currently no formal definitions for estimands which encompass both concepts, nor any guidance on how these attributes differ for the construction of estimands. Finally, because these estimands have only been considered separately, the literature around estimation has also typically focussed on only a single attribute of the estimand (e.g. estimation of a marginal effect, or estimation of a participant-average effect), meaning there is currently no guidance on estimation of estimands which incorporate both the marginal vs. cluster-specific and participant- vs. cluster-average attributes unique to CRTs.

The purpose of this paper is therefore to resolve these issues by (i) defining estimands which incorporate both attributes together (e.g. marginal, participant-average estimands; cluster-specific, participant-average estimands; etc), and demystifying these interconnected concepts and terminology; and (ii) describing estimators that can be used for each of these estimands.

## 2. Estimands

In this section we describe the difference between marginal and cluster-specific estimands (section 2.2), then separately we describe the difference between participant-average and cluster-average estimands (noting when the different estimands will coincide) (section 2.3). A summary is provided in Table 1.

We then merge the two concepts to define estimands incorporating both attributes; because there are two options for each concept, this leads to four total estimands: (i) marginal, participant-average; (ii) marginal, cluster-average; (iii) cluster-specific, participant-average; and (iv) cluster-specific, cluster average. These are defined for both a difference in means or proportions and an odds ratio in section 2.4, and Table 2.

We note that to be fully defined, each of the estimands described below would require specification of the other attributes encompassing an estimand (i.e. population, treatment conditions, endpoint, summary measure, and handling of intercurrent events [1, 30]). Some of these attributes may require additional consideration in CRTs; for instance, the population of interest would need to be described for clusters as well as participants, and investigators may need to differentiate whether interest lies in the population of all eligible participants vs. just those that would enrol in the study if provided the opportunity [31, 32]. Similarly, in some CRTs, investigators may need to describe the duration of the implementation of the intervention (e.g. the average effect over three vs. six months of implementation [33, 34]) when describing the treatment conditions of interest. Similarly, because non-adherence could occur both at the participant or cluster level, they may need to define intercurrent events at both the individual and cluster-level. To our knowledge, there are unlikely to be any additional consideration in CRTs when defining the endpoint or summary measure attributes (apart from the cluster-specific vs. marginal distinction discussed in this paper). To facilitate a clearer description of our main message, we do not further address these additional considerations below.

We describe each estimand using the potential outcomes framework. We do this under two different perspectives that have been used for defining causal effects in CRTs: (i) a super-population perspective, where each cluster is assumed to be an independent random sample from a hypothetical infinite population of clusters [26, 29, 35]; and (ii) a finite-population perspective, where the clusters in the trial are themselves considered as the target population [28]. The key difference between these is that under the super-population perspective the estimand is written in terms of a population expectation (taken over the infinite population of clusters), while under the finite-population perspective the estimand is written as an empirical average across the observed clusters and participants in the study. Though the concepts behind the participant- vs. cluster-average and marginal vs. cluster-specific aspects are the same under each perspective, we provide both for completeness. For simplicity however, we focus on describing the estimands under the finite-population perspective in what follows and include those under the super-population perspective in Table 1.

### 2.1 Notation

We begin by introducing the notation that will be used to define both the estimands and the estimators below. Let $Y_{ij}$ denote the observed outcome for participant $i$ from cluster $j$, and $Z_j$ represent the treatment assignment for cluster $j$. Further, let $n_j$ be the number of participants in cluster $j$, and $M$ and $N$ represent the total number of clusters and the total number of participants respectively.

Under the potential outcome framework, $Y_{ij}^{(1)}$ denotes the potential outcome that would have been realized under $Z_j = 1$ for participant $i$ from cluster $j$, and similarly $Y_{ij}^{(0)}$ represents that participant's potential outcome under $Z_j = 0$. Then, let $\bar{Y}_j^{(1)}$ denote the average potential outcome in cluster $j$ under treatment $Z_j = 1$ i.e.:

$$\bar{Y}_j^{(1)} = \frac{1}{n_j} \sum_{i=1}^{n_j} Y_{ij}^{(1)} \quad (1)$$

and similarly for $\bar{Y}_j^{(0)}$ under the control condition.

## 2.2 Marginal and cluster-specific estimands

We provide formal definitions for marginal and cluster-specific estimands below. Briefly, the difference between marginal and cluster-specific estimands is based on how the potential outcomes are summarised.

A marginal estimand is calculated using the following steps:

1) Overall summaries are obtained by summarising the potential outcomes separately by treatment condition using *all* participants (e.g. the mean potential outcome is calculated under the intervention and control respectively);
2) The summaries are contrasted between treatment conditions to calculate an overall marginal treatment effect.

Conversely, for a cluster-specific estimand, the following steps are taken:

1) Cluster-specific summaries are obtained by summarising the potential outcomes *within each cluster* (e.g. the mean potential outcome is calculated under intervention and control respectively in cluster 1, cluster 2, etc);
2) The cluster-specific summaries are contrasted between treatment conditions within each cluster to calculate a cluster-specific treatment effect;
3) An average of these cluster-specific effects is calculated to provide the overall cluster-specific estimand (note this average can be taken in different ways; see section 2.3. on participant- vs. cluster-average estimands below).

Thus, the difference between marginal and cluster-specific estimands is whether an overall summary measure is calculated within each treatment arm before the arms are contrasted (marginal estimand), or whether summary measures are contrasted within each cluster first (cluster-specific estimand).

For certain summary measures, the overall cluster-specific estimand may average over some function of the cluster-specific effects. For instance, when defining an estimand based on an odds ratio, the average of log odds ratio may be taken across clusters, then back-transformed after to obtain the overall cluster-specific odds ratio. This example will be further discussed in section 2.4.

We note that marginal and cluster-specific estimands can be written as either participant-average or cluster-average treatment effects (depending on how each individual or cluster will be weighted, described in section 2.3); for simplicity, we describe the differences between marginal and cluster-specific estimands below in terms of participant-average effects.

### 2.2.1 Marginal estimands

For a difference in means or proportions, a marginal participant-average estimand is defined under the finite-population perspective as:

$$\Delta^{MG-PA} = \frac{1}{N}\sum_{j=1}^{M}\sum_{i=1}^{n_j} Y_{ij}^{(1)} - \frac{1}{N}\sum_{j=1}^{M}\sum_{i=1}^{n_j} Y_{ij}^{(0)} \quad (2)$$

### 2.2.2 Cluster-specific estimands

The cluster-specific estimand for a difference in means/proportions can be defined under the finite-population perspective as follows. First, let $\beta_j$ represent the cluster-specific treatment effect for cluster $j$, i.e.:

$$\beta_j = \bar{Y}_j^{(1)} - \bar{Y}_j^{(0)} \quad (3)$$

where $\bar{Y}_j^{(1)}$ and $\bar{Y}_j^{(0)}$ are defined based on equation (1). Then, the cluster-specific participant-average estimand is defined based on a weighted linear combination of the $\beta_j$'s:

$$\Delta^{CS-PA} = \frac{\sum_{j=1}^{M} n_j \beta_j}{\sum_{j=1}^{M} n_j} = \frac{1}{N}\sum_{j=1}^{M}\sum_{i=1}^{n_j} Y_{ij}^{(1)} - \frac{1}{N}\sum_{j=1}^{M}\sum_{i=1}^{n_j} Y_{ij}^{(0)} \quad (4)$$

### 2.2.3 When will marginal and cluster-specific estimands coincide or differ?

We use the term *collapsible* to indicate that the values of the two estimands will coincide, and *non-collapsible* to mean that the values of the two estimands will differ [36].

Whether the marginal and cluster-specific estimands will coincide or not depends on the summary measure (e.g. difference, odds ratio) being used. For differences (e.g. difference in means, difference in proportions), these two estimands will coincide (i.e., $\Delta^{MG-PA} = \Delta^{CS-PA}$) because the empirical average is a linear operator. Specifically, we can see from equation (4) that it is mathematically equivalent to first summarise outcomes overall with an empirical average and then take a difference, or to take a difference within clusters and then take an average of these differences. Because the two estimands are the same, a "difference" summary measure is *collapsible*.

For ratio summary measures (e.g. risk ratios, odds ratios), in general the two estimands will differ. This is because of the function used which transforms the summaries (e.g. by taking the log or logit transformation of the marginal or cluster-specific summaries); this feature renders the mathematical equivalency stated above invalid; that is, a ratio of overall summaries is generally *not* the same as an average of the ratios within each cluster (except in a few specific settings, e.g. if the risk ratio is identical across all clusters). Because the marginal and cluster-specific estimands are different, "ratio" summary measures are *non-collapsible*, except in special circumstances.

## 2.3 Participant-average and cluster-average estimands

We provide formal definitions for participant- and cluster-average estimands under the finite-population perspective below. Briefly, the difference between the participant- and cluster-average estimands is in how the potential outcomes are weighted [4]. Under the participant-average estimand, a general principle is that each participant is given equal weight. Under the cluster-average effect definition, a general principle is that each cluster is given equal weight (implying that participants from smaller clusters are given more weight than participants from larger clusters) [4].

We note that both the participant- and cluster-average treatment effects can be written as either marginal or cluster-specific estimands; however, in this section, we write them as marginal estimands for simplicity.

### 2.3.1 Participant-average estimands

For a difference in means, or difference in proportions, the marginal participant-average estimand is given in equation (2) above (as we defined marginal estimands above as participant-average for simplicity). We repeat this equation here for completeness:

$$\Delta^{MG-PA} = \frac{1}{N}\sum_{j=1}^{M}\sum_{i=1}^{n_j} Y_{ij}^{(1)} - \frac{1}{N}\sum_{j=1}^{M}\sum_{i=1}^{n_j} Y_{ij}^{(0)}$$

To provide additional insights into this definition, an alternative representation of the marginal participant-average estimand is [28]:

$$\Delta^{MG-PA} = \frac{1}{M}\sum_{j=1}^{M}\frac{n_j M}{N}\bar{Y}_j^{(1)} - \frac{1}{M}\sum_{j=1}^{M}\frac{n_j M}{N}\bar{Y}_j^{(0)}$$

Which reveals that the weight to each cluster-specific summary measure ($\bar{Y}_j^{(1)}$ and $\bar{Y}_j^{(0)}$, which are defined based on equation (1)) is proportional to its cluster size $n_j$, implying that a larger cluster is given more weight than a smaller cluster. This representation is helpful as it as it underlies a key distinction between participant- and cluster-average estimands.

### 2.3.2 Cluster-average estimands

For a difference in means/proportions, the marginal cluster-average estimand is defined as:

$$\Delta^{MG-CA} = \frac{1}{M}\sum_{j=1}^{M}\bar{Y}_j^{(1)} - \frac{1}{M}\sum_{j=1}^{M}\bar{Y}_j^{(0)} \quad (5)$$

As above, to provide additional insights into this definition, we provide an alternative representation of the marginal cluster-average estimand:

$$\Delta^{MG-CA} = \frac{1}{N}\sum_{j=1}^{M}\left(\frac{1}{n_j}\sum_{i=1}^{n_j}Y_{ij}^{(1)}\right) - \frac{1}{N}\sum_{j=1}^{M}\left(\frac{1}{n_j}\sum_{i=1}^{n_j}Y_{ij}^{(0)}\right)$$

This clearly demonstrates that cluster-average estimands gives equal weight to each cluster, regardless of the cluster size used to generate the summaries $\bar{Y}_j^{(1)}$ and $\bar{Y}_j^{(0)}$.

### 2.3.3 When will participant- and cluster-average estimands coincide or differ?

Whether the participant- and cluster-average estimands will differ depends on two things: (i) whether the estimand summary measure being used is *collapsible* (e.g. a difference in means or proportions) or *non-collapsible* (e.g. a ratio, except in specific circumstances); and (ii) whether informative cluster size is present. Informative cluster size occurs when either the potential outcomes or potential outcome contrasts (that is, $Y_{ij}^{(1)} - Y_{ij}^{(0)}$ for a difference in means or proportions) depends on cluster size [37, 38].

For collapsible summary measures, such as the difference in means/proportions, the participant- and cluster-average estimands will differ when the second type of informative cluster size is present, i.e. when the potential outcome contrasts differs according to cluster size [4].

For non-collapsible summary measures, such as the odds ratio, the participant- and cluster-average treatment effect will typically differ when either the first or second type of informative cluster size is present, i.e. when either the outcomes or the treatment effects differ depending on the cluster size [4].

## 2.4 Estimands encompassing both attributes (marginal vs. cluster-specific, participant- vs. cluster-average)

Estimand definitions incorporating both attributes (marginal vs. cluster-specific attribute, and participant- vs. cluster-average attribute) are described below, and summarised in Table 2. Because there are two options for each attribute, this leads to four total estimands. Below we describe the construction of these estimands for both a difference and an odds ratio summary measure.

### 2.4.1 Marginal, participant-average

The marginal participant-average estimand for a difference is given in equation (2) above, and we repeat it here for convenience:

$$\Delta^{MG-P} = \frac{1}{N}\sum_{j=1}^{M}\sum_{i=1}^{n_j} Y_{ij}^{(1)} - \frac{1}{N}\sum_{j=1}^{M}\sum_{i=1}^{n_j} Y_{ij}^{(0)}$$

Similarly, the marginal participant-average odds ratio (we use the notation Γ to denote the odds ratio estimands to differentiate from the difference-in-means notation Δ) for binary outcomes can be defined as:

$$\Gamma^{MG-PA} = \frac{\frac{1}{N}\sum_{j=1}^{M}\sum_{i=1}^{n_j} Y_{ij}^{(1)} / \left(1 - \frac{1}{N}\sum_{j=1}^{M}\sum_{i=1}^{n_j} Y_{ij}^{(1)}\right)}{\frac{1}{N}\sum_{j=1}^{M}\sum_{i=1}^{n_j} Y_{ij}^{(0)} / \left(1 - \frac{1}{N}\sum_{j=1}^{M}\sum_{i=1}^{n_j} Y_{ij}^{(0)}\right)} \quad (6)$$

### 2.4.2 Marginal, cluster-average

For a difference in means/proportions, the marginal cluster-average estimand is given in equation (5); we repeat it here for convenience:

$$\Delta^{MG-CA} = \frac{1}{M}\sum_{j=1}^{M} \bar{Y}_j^{(1)} - \frac{1}{M}\sum_{j=1}^{M} \bar{Y}_j^{(0)}$$

And the marginal cluster-average odds ratio for binary outcomes is:

$$\Gamma^{MG-C} = \frac{\frac{1}{M}\sum_{j=1}^{M} \bar{Y}_j^{(1)} / \left(1 - \frac{1}{M}\sum_{j=1}^{M} \bar{Y}_j^{(1)}\right)}{\frac{1}{M}\sum_{j=1}^{M} \bar{Y}_j^{(0)} / \left(1 - \frac{1}{M}\sum_{j=1}^{M} \bar{Y}_j^{(0)}\right)} \quad (7)$$

where definitions of $\bar{Y}_j^{(1)}$ and $\bar{Y}_j^{(0)}$ are given in equation (1). In other words, this estimand is constructed by applying the odds ratio summary measure to the cluster-average potential outcomes, $\frac{1}{M}\sum_{j=1}^{M} \bar{Y}_j^{(1)}$ and $\frac{1}{M}\sum_{j=1}^{M} \bar{Y}_j^{(0)}$.

### 2.4.3 Cluster-specific, participant-average

The cluster-specific participant-average estimand for a difference in means/proportions is given in equation (4); we repeat it here for convenience:

$$\Delta^{CS-P} = \frac{\sum_{j=1}^{M} n_j \beta_j}{\sum_{j=1}^{M} n_j}$$

where $\beta_j$ was defined in equation (3). For a difference-in-means/proportions, $\Delta^{CS-PA} = \Delta^{MG-PA}$ due to collapsibility.

For the cluster-specific participant-average odds ratio, one can first let $OR_j$ denote the cluster-specific odds ratio for cluster $j$, i.e.:

$$OR_j = \frac{\frac{1}{n_j}\sum_{i=1}^{n_j} Y_{ij}^{(1)} / \left(1 - \frac{1}{n_j}\sum_{i=1}^{n_j} Y_{ij}^{(1)}\right)}{\frac{1}{n_j}\sum_{i=1}^{n_j} Y_{ij}^{(0)} / \left(1 - \frac{1}{n_j}\sum_{i=1}^{n_j} Y_{ij}^{(0)}\right)} \quad (8)$$

Then, the cluster-specific participant-average effect can be defined based on a cluster size weighted average as:

$$\frac{\sum_{j=1}^{M} n_j f(OR_j)}{\sum_{j=1}^{M} n_j} \quad (9)$$

where $f(OR_j)$ is some function of the cluster-specific odds ratio, for instance, the identity or log function. Each function corresponds to a different way of averaging the cluster-specific odds ratios across clusters. Typically, a log function, where $f(OR_j) = log(OR_j)$, would be most natural for odds ratios. This is because (1) the log odds ratio is connected to regression parameters with a canonical link function and is a familiar concept, and (2) the resulting average in equation (9) can be interpreted as a cluster size weighted geometric mean of $OR_j$, i.e.,

$$log\left\{\left(\prod_{j=1}^{M} OR_j^{n_j}\right)^{\frac{1}{\sum_{j=1}^{M} n_j}}\right\}.$$

Importantly, when a function other than identity is used, the resulting average may need to be back transformed to the appropriate scale. For example, when $f$ is a log function, a back transformation should be applied to equation (9) to obtain the final cluster-specific, participant-average estimand, that is,

$$\Gamma^{CS-PA} = \exp\left(\frac{\sum_{j=1}^{M} n_j \log(OR_j)}{\sum_{j=1}^{M} n_j}\right)$$

Of note, the estimand in (9) can be seen to correspond to a participant-average effect because each participant gets equal weight in the construction of the cluster-specific contrasts of potential outcomes (the $OR_j$'s in equation (8)), and each cluster-specific contrast is weighted according to the number of participants belonging to that cluster.

In general, cluster-specific estimands (participant- or cluster-average) are only well defined when the cluster-specific treatment effects are well defined, e.g. for an odds ratio this would require the potential outcome proportion to be bounded away from 0 or 1 in each cluster so that a cluster-specific ORs from equation (8) can be defined without ambiguity.

2.4.4 Cluster-specific, cluster-average

The cluster-specific cluster-average estimand for a difference in means/proportions is defined as:

$$\Delta^{CS-CA} = \frac{1}{M}\sum_{j=1}^{M} \beta_j \quad (10)$$

where $\beta_j$ was defined in equation (3). For a difference-in-means, $\Delta^{CS-CA} = \Delta^{MG-C}$ due to collapsibility.

The cluster-specific cluster-average odds ratio can be defined by giving equal weight to each cluster-specific summary:

$$\frac{1}{M}\sum_{j=1}^{M} f(OR_j) \quad (11)$$

Where $OR_j$ was defined in equation (8), and $f(OR_j)$ was defined as in section 2.4.3. When $f$ is a log function, equation (11) is interpreted as the geometric mean of $OR_j$ as $log\left\{\left(\prod_{j=1}^{M} OR_j\right)^{\frac{1}{M}}\right\}$. In this case, a final back transformation should be applied to equation (11) to define the final cluster-specific, cluster-average estimand as:

$$\Gamma^{CS-C} = \exp\left(\frac{1}{M}\sum_{j=1}^{M} log(OR_j)\right)$$

## 3. Estimators

We now describe some familiar estimators that can be used to estimate each of the estimands described previously for a parallel arm CRT. We focus on the following estimators: (a) independence estimating equations (IEEs) [4, 19]; (b) the analysis of cluster-level summaries [4, 15, 18]; (c) mixed-effects models with a cluster-level random intercept; and generalised estimating equations (GEEs) with an exchangeable working correlation structure. We focus on the simple scenario without baseline covariate adjustment for each estimand. When covariate adjustment is of interest to obtain more efficient estimators, we refer readers to Su and Ding [28] for related development under a finite-population perspective, Balzer *et al.*, Benitez *et al* and Wang *et al.* [25, 29, 39, 40] for related developments under a super-population perspective; although these prior developments often focus only on a subset of estimands we have covered. Below, we address commonly used estimators for CRTs in the absence of covariate adjustment, and then explain which estimands they target, along with the key assumptions required for consistency (i.e. asymptotically unbiased estimation). A summary is given in Table 3.

In general, the assumptions required to consistently estimate the target estimand are similar under both perspectives (super population versus finite-population), except for some technical differences in conceptualizing the asymptotic regime and versions of the Central Limit Theorems invoked [41, 42]; for simplicity, we do not distinguish between these two perspectives and only discuss the necessary assumptions with easy-to-understand terms.

We note that this list of estimators we describe is not intended to be comprehensive. In particular, we focus on the standard implementations of mixed-effects models and GEEs with an exchangeable correlation structure, in which the estimated treatment effect is taken from the model parameter corresponding to the assigned treatment. However, there are other implementations that could be used (e.g., model-based g-computation estimators based on linear-mixed models and GEEs in Section 3 in Wang *et al* [29], and propensity score weighting estimators in Zhu *et al* [43]) which can consistently estimate the participant-average and cluster-average estimands even when the associated working models are misspecified.

In general, all estimators described above require the following assumptions: (i) the consistency assumption (sometimes termed the cluster-level stable unit treatment value assumption), which requires that $Y_{ij} = Y_{ij}^{(z)}$ for $Z = z$, i.e. that a participant's observed outcome is equal to their potential outcome under their allocated treatment and is defined without ambiguity [44]; (ii) the maximum cluster size is bounded; (iii) exchangeability between treatment arms [44] (i.e. that clusters are randomised between treatment arms, and there is no differential enrolment of patients between treatments [31, 32]); (iv) observations are independent across clusters [29]; (v) a large number of clusters such that appropriate versions of Law of Large Numbers and Central Limit Theorems can be applied.

In addition, estimators for cluster-specific estimands with a binary outcome will also require the assumption that the average potential outcome in each cluster is bounded away from 0 or 1. We discuss additional assumptions required for mixed-effects models and GEEs with an exchangeable correlation structure below.

Of note, we slightly abuse the notation throughout this Section such that $\beta$ represents the treatment regression coefficient across different working models. We caution that the regression coefficient should be interpreted based on each working models separately and not compared across models.

### 3.1 Independence estimating equations (IEEs)

Independence estimating equations (IEEs) are a class of estimators which is applied to individual participant outcomes and uses an independence working correlation structure in conjunction with cluster-robust standard errors (SEs) [45]. Briefly, IEEs make a working assumption that outcomes within a cluster are independent; in practice, this assumption will almost always be false for CRTs however it helps ensure consistent (asymptotically unbiased) estimation in the presence of informative cluster size [4]. The cluster-robust SEs then serve to ensure estimated standard errors are asymptotically valid despite the incorrect working independence assumption [46].

IEEs can be used to estimate marginal, participant-average effects, as well as marginal, cluster-average effects (using different specifications of weights). However, they cannot be used to estimate cluster-specific effects. They do not require any assumptions beyond those specified earlier.

### 3.1.1 Marginal, participant-average estimator

For a difference in means, IEEs can be implemented to estimate the marginal, participant-average effect by applying the following model to individual participant data:

$$Y_{ij} = \alpha + \beta Z_j + \varepsilon_{ij} \quad (12)$$

In this model, an independence working correlation structure and a constant variance structure are specified for $\varepsilon_{ij}$. Estimation of $\beta$ can then be done either using a linear regression model (which makes the working independence assumption automatically), or by using generalised estimating equations (GEEs) with a working independence correlation structure alongside an "*identity*" link and "*Gaussian*" family. Importantly, for both approaches (linear regression, GEEs) cluster-robust SEs must be specified to account for correlation between outcomes within the same cluster [28, 45].

Similarly, the marginal, participant-average odds ratio can be estimated through the following model:

$$logit(P\{Y_{ij} = 1\}) = \alpha + \beta Z_j \quad (13)$$

As above, this model also uses an independence working correlation structure, and could be applied either using a standard logistic regression model or GEEs with a working independence correlation structure alongside a "*logit*" link and "*binomial*" family (making sure to use cluster-robust SEs in both cases) [43]. Of note, to estimate the final marginal, participant-average odds ratio, one need to exponentiate the treatment effect coefficient in (13), that is, $\Gamma^{MG-} = \exp(\beta)$; a proof of consistency under the super-population perspective can be found in Web Appendix 1 of Zhu *et al* [43].

Because the models in equations (12) and (13) give equal weight to each participant, they correspond to a participant-average effect, and because the models first summarise outcomes within treatment groups before applying any transformations, they correspond to marginal effects.

### 3.1.2 Marginal, cluster-average estimator

IEEs can be used to estimate a marginal, cluster-average effect using models (12) and (13) above, however each individual observation is additionally weighted by the inverse cluster size $\frac{1}{n_j}$. This is to ensure each cluster is given an equal weight of 1 (i.e. weighting by $\frac{1}{n_j}$ ensures the sum of weights across participants in each cluster is equal to 1). As above, an independence working correlation structure is used alongside cluster-robust SEs.

Because these models give equal weight to each cluster, they correspond to a cluster-average effect. As above, because they summarise outcomes within treatment groups before applying any transformations, they correspond to marginal effects. In the case of an odds ratio summary measure, a simple modification of the proof in Web Appendix 1 of Zhu *et al.* can be used to show that this weighted IEE estimator is consistent [43].

### 3.2 Analysis of cluster-level summaries

The analysis of cluster-level summaries involves two steps: (i) a summary measure is taken in each cluster (e.g. the mean observed outcome across all participants in the cluster); and (ii) the analysis is performed on the cluster-level summaries.

The analysis of cluster-level summaries can be used to estimate all four estimands described previously (marginal, participant-average effects; marginal, cluster-average effects; cluster-specific participant-average effects; and cluster-specific, cluster-average effects). For illustration, we describe the different implementations below for an odds ratio summary measure (we note that for a difference in means/proportions, implementations of marginal and cluster-specific estimators are identical; this is because no transformation of the marginal/cluster-specific summaries needs to be taken). They require the standard assumptions specified earlier. Further, the cluster-specific estimators require the same assumption as required for the cluster-specific estimand, that is, that the potential outcome proportions need to be bounded away from 0 or 1 in each cluster so that an odds ratio summary within each cluster is well-defined. We provide proof of consistency (i.e. asymptotically unbiased estimation) for each cluster-level summary approach described below in section 7.

### 3.2.1 Marginal, participant-average estimator

This estimator uses a two-step procedure. In the first step, the proportion of events in each cluster is calculated, represented by $\hat{\pi}_j$ for cluster $j$. Then, a weighted generalised linear model (GLM) using an appropriate link function (logit for an odds ratio) is applied using the $\hat{\pi}_j$'s as outcomes, i.e.:

$$logit(\hat{\pi}_j) = \alpha + \beta Z_j \quad (14)$$

and during estimation each cluster-level summary is weighted by $n_j$ (to give equal weight to each participant). To implement the GLM, a working distribution family must also be chosen (e.g. binomial, Gaussian, etc). We show in Section 7.1 different choices of distribution family for the model in (14) has no impact on the estimation of $\beta$; this is a special result since the model has no additional covariates beyond the treatment indicator. For simplicity, we use of a "*Gaussian*" family which is consistent with the standard implementation of a cluster-level summaries approach (in which the summaries are compared in a linear regression model; this is described below). The final odds ratio parameter is calculated by $\Gamma^{MG-PA} = exp\{\beta\}$.

Because this model gives equal weight to each participant, it corresponds to a participant-average effect. Furthermore, because this model summarises outcomes within treatment group before applying any transformations, it estimates a marginal effect.

### 3.2.2 Marginal, cluster-average estimator

Cluster-level summaries can be used to estimate a marginal, cluster-average effect, using model (14) above, however the cluster-level summaries are unweighted in order to give equal weight to each cluster. The final odds ratio parameter is calculated by $\Gamma^{MG-CA} = exp\{\beta\}$. As above, in addition to showing the consistency of this estimator, in Section 7.2 we demonstrate that the choice of distribution family has no impact on the estimation of $\beta$.

### 3.2.3 Cluster-specific, participant-average estimator

This estimator uses a three-step procedure, as follows:

1. . As above, the proportion of observed events in each cluster ($\hat{\pi}_j$) is calculated.
2. The proportions from step 1 are transformed according to the summary measure being used; for instance, an odds ratio would require calculating the log-odds in each cluster, i.e. $\log(odds_j) = \log(\hat{\pi}_j/\{1 - \hat{\pi}_j\})$.
3. Finally, the cluster-level summaries calculated in step 2 (e.g. the $\log(odds_j) = \log(\hat{\pi}_j/\{1 - \hat{\pi}_j\})$) are analysed using model (15) below, which is a weighted linear regression working model (where each cluster-level summary weighted by $n_j$, and the treatment indicator is the independent variable).

$$\log(odds_j) = \alpha + \beta Z_j + \varepsilon_j \quad (15)$$

where $\log(odds_j)$ was defined in step 2. Note that this model could be equivalently written as:

$$logit(\hat{\pi}_j) = \alpha + \beta Z_j + \varepsilon_j$$

The treatment effect estimate is then back transformed as appropriate (e.g. the odds ratio is then calculated as $\Gamma^{CS-PA} = exp\{\beta\}$). As discussed above, for a difference in means/proportions, no transformation is required (i.e. the cluster-specific and marginal estimators are identical).

Because this estimation procedure weights each cluster-specific summary by the cluster size (hence intuitively giving each participant the same weight), it corresponds to a participant-average effect. Furthermore, because it applies transformations to the cluster-level summaries directly (rather than summarising the entire treatment arm before applying the transformation) it targets a cluster-specific effect. As above, we sketch the proof for consistency in Section 7.3.

### 3.2.4 Cluster-specific, cluster-average estimator

Cluster-level summaries can be used to estimate a cluster-specific, cluster-average effect, using model (15) above, however the cluster-level summaries are unweighted in order to give equal weight to each cluster. A proof of consistency is given in Section 7.4.

### 3.3 Mixed-effects models

Mixed-effects models are applied to participant level data and involve specifying a random intercept term for the clusters. For a difference in means, a linear mixed-effects model takes the form:

$$Y_{ij} = \alpha + \beta Z_j + \mu_j + \varepsilon_{ij} \quad (16)$$

where $\mu_j$ represents a random effect for cluster $j$ which is assumed to follow a normal distribution with mean 0 and variance $\sigma_B^2$, and $\varepsilon_{ij}$ is a random error term for participant $i$ from cluster $j$ which is assumed to follow a normal distribution with mean 0 and variance $\sigma_W^2$. Estimation is performed using maximum likelihood (or restricted maximum likelihood).

For estimating an odds ratio, a logistic mixed-effects model with a random intercept is:

$$logit(P\{Y_{ij} = 1\}) = \alpha + \beta Z_j + \mu_j \quad (17)$$

Mixed-effects models have been conventionally considered as tools to estimate a cluster-specific, participant-average effect. However, by construction of the estimators, they do not in fact give equal weight to each participant (or, equivalently, weight each cluster by its respective cluster size). In fact, Wang et al. [19] have shown that the generalized least squares estimator of $\beta$ in model (16) targets the following quantity (expressed under a finite-population perspective):

$$\Delta(\rho^*) = \frac{\sum_{j=1}^{M} \frac{n_j}{1 + (n_j - 1)\rho^*} \left\{ \bar{Y}_j^{(1)} - \bar{Y}_j^{(0)} \right\}}{\sum_{j=1}^{M} \frac{n_j}{1 + (n_j - 1)\rho^*}}$$

where $\rho^*$ is the probability limit of the intracluster correlation estimator, $\hat{\sigma}_B^2 / (\hat{\sigma}_B^2 + \hat{\sigma}_W^2)$. Thus, clusters are weighted by their inverse-variance (also referred to as the precision weights), with weights additionally depending on the unknown variance components [4]. While this weighting scheme is motivated by efficiency consideration, it lacks the ready interpretability of cluster- or participant-average approaches. Furthermore it is not an appropriate estimand as it depends on an unknown parameter $\rho^*$ which is specific to particular contexts or datasets. For instance, if we analyse a different outcome (switching from a continuous outcome to a binary outcome), we would expect a different intracluster correlation value $\rho^*$, which would lead to a re-weighting of the participants and clusters in constructing $\Delta(\rho^*)$. This would not be the case for the other estimands defined earlier as those weighting schemes are independent of the outcome used. Hence, choosing $\Delta(\rho^*)$ as an estimand implicitly means that the estimand will also be dictated by the data we analyse rather than the scientific question alone. Under informative cluster size, this estimand generally differs from the estimands we defined in Section 2.4, except for two extreme cases. That is, when $\rho^* = 0$, we have $\Delta(\rho^*) = \Delta^{CS-} = \Delta^{MG-PA}$, and when $\rho^* = 1$, we have $\Delta(\rho^*) = \Delta^{CS-C} = \Delta^{MG-C}$. However, these two cases are generally unlikely to hold for most CRTs. As such, mixed-models may generally be biased for the cluster-specific, participant average effect when there is informative cluster size. Of note, they will also be biased for other estimands as well, such as the cluster-specific, cluster-average effect, when there is informative cluster size.

However, in the absence of informative cluster size, linear mixed-effects models can provide a consistent estimator for "difference" summary measure for all four estimands, as in this case, the values of all estimands, $\{\Delta(\rho^*), \Delta^{MG-P}, \Delta^{MG-CA}, \Delta^{CS-PA}, \Delta^{CS-C}\}$, will coincide (strictly speaking, they all coincide under the super-population perspective, whereas their differences vanish with $M \to \infty$ under the finite-population perspective). Wang et al. provides a detailed treatment on the robustness of linear mixed-effects models for CRTs under arbitrary model misspecifications in the absence of informative cluster size [47].

However, these results do not easily generalize to other link functions such as the logistic mixed-effects models, since the marginal likelihood and the score equations are analytically intractable. Therefore, the requirement for consistent estimation of the odds ratio estimands (defined in Section 2) with logistic mixed-effects models is likely more stringent compared to linear mixed-effects models. To this end, an important area of future research is around whether deviations from model assumptions (e.g., that the normality assumption for the random-effects in the logistics mixed-effects model is misspecified) may affect the consistent estimation of all four odds ratio estimands.

3.4 Generalised estimating equations with an exchangeable correlation structure
GEEs are applied to individual participant data [45]. They involve specifying a working correlation structure in conjunction with cluster-robust SEs. Typically, for CRTs an exchangeable working correlation structure is specified, i.e. the correlation is assumed to be constant between any two participants in the same cluster, and 0 between participants from different clusters. The use of the cluster-robust SE ensures consistent variance estimation, even when the working correlation structure is misspecified. For a difference-in-means, GEEs take the following mean model:

$$E(Y_{ij}) = \alpha + \beta Z_j \quad (18)$$

and a working correlation structure is specified for $\left(Y_{1j}, \dots, Y_{n_j,j}\right)'$ for each cluster. Estimation is done using a pair of estimating equations, one for the mean parameters and one for the correlation parameters [45].

For an odds ratio summary measure, the following form is used:

$$logit(P\{Y_{ij} = 1\}) = \alpha + \beta Z_j \quad (19)$$

GEEs with an exchangeable correlation structure have been conventionally considered as tools to estimate a marginal, participant-average effect. However, in the presence of informative cluster size, just like mixed-effects models, they do not in fact give equal weight to each participant, and as such, they require non-informative cluster size to provide consistent estimation of estimands defined earlier [19]. See, for example, Wang *et al.* for a detailed overview on the robustness of GEEs with an exchangeable correlation structure and an identity link function [19].

## 4. Application to the RESTORE trial

The Randomized Evaluation of Sedation Titration for Respiratory Failure (RESTORE) trial was a CRT that compared protocolised sedation with usual care in critically ill children who were mechanically ventilated for acute respiratory failure [48]. Thirty-one clusters were randomised, with the number of participants in each cluster ranging between 12 to 272. In total, 2449 participants were enrolled.

We focus on the adverse event "postextubation stridor", which denoted the presence of inspiratory noise indicating the narrowing of the airway (yes vs. no). Our aims were (i) to compare estimators for the same estimand, to determine to what extent different choices may impact results; and (ii) to compare estimators across different estimands, to evaluate to what extent choice of estimand may affect interpretation of trial results.

We implemented each of the estimators described in section 3. For IEEs and GEEs we calculated cluster-robust SEs using the Fay-Graubard small-sample correction [49]. For the analysis of cluster-level summaries we used Huber-White SEs, and for mixed-effects models we used model-based SEs.

### 4.1 Difference between estimators

Results are shown in Table 4 and Figure 1. The estimated odds ratio ranged substantially across different estimands, from 1.38 (95% CI 0.87 to 2.19, p=0.17) for the marginal cluster-average effect, to 1.83 (95% CI 1.06 to 3.14, p=0.03) for the cluster-specific participant-average effect.

The choice of both estimand and estimator impacted on conclusions. Estimates for the participant-average estimands (both marginal and cluster-specific) were larger than those for the cluster-average estimands, and were statistically significant (based on a 0.05 significance level), while those for the cluster-average estimands were not.

However, only specific estimators for the two participant-average estimands demonstrated statistical significance. In particular, mixed-effects models and GEEs with an exchangeable correlation structure produced smaller estimates of treatment effect than IEEs or the analysis of cluster-level summaries, and, as a consequence, results from mixed-effects models and GEEs with an exchangeable correlation structure were not statistically significant, while those from IEEs and cluster-level summaries were. For example, for the marginal participant-average effect, IEEs provided an estimated odds ratio (OR) of 1.65 (95% CI 1.02 to 2.67, p=0.04), while GEEs with an exchangeable correlation structure produced an estimated OR of 1.57 (95% CI 0.98 to 2.50, p=0.06). Similarly, for the cluster-specific participant-average effect, the use of weighted cluster-level summaries provided an OR of 1.83 (95% CI 1.06 to 3.14, p=0.03), while a mixed-effects logistic regression model gave an estimated OR of 1.54 (95% CI 0.97 to 2.44, p=0.07).

The participant-average effects were larger than the corresponding cluster-average effects, which is consistent with the implication of informative cluster size [50]. Smaller clusters had numerically smaller treatment effects than larger clusters: the participant-average OR was 1.15 (95% CI 0.56 to

2.36) in the 24 clusters of size <100, while it was 2.26 (95% CI 1.49 to 3.43) in the 7 clusters of size ≥100. We also observed attenuated estimates from mixed-effects models and GEEs with an exchangeable correlation structure. Although often thought to estimate participant-average effects, in fact these models can give more weight to smaller clusters. Hence, they may give an estimate 'shifted' towards the cluster-average effect, in this case a smaller overall treatment effect [4]. However, the interaction between small and large clusters was not statistically significant (p=0.21), so we cannot definitively conclude there was informative cluster size in this setting.

## 5. Discussion

The use of estimands to clarify the interpretation of treatment effects and ensure that estimators are aligned with study objectives has rapidly been gaining attention in randomised trials, however most research has been focussed on individually randomised trials. CRTs have unique features that require additional specification in the estimand definition. In this paper we have: (i) defined estimands that encompass both the marginal vs. cluster-specific and participant- vs. cluster-average attributes together; and (ii) described commonly used, simple estimators for each of these estimands.

Our re-analysis of the RESTORE trial demonstrated the value of careful consideration of both the estimand and the estimator. Different estimands led to different conclusions around the effect of treatment (e.g. OR=1.38, p=0.17 for the marginal cluster-average effect vs. OR=1.83, p=0.03 for the cluster-specific participant-average effect). Similarly, different estimators of the same estimand also affected interpretation. Mixed-effects models and GEEs with an exchangeable correlation structure, which may be considered for estimation of participant-average effects (cluster-specific and marginal respectively), provided lower estimates of treatment effect that were closer to the cluster-average effect than either IEEs or the analysis of cluster-level summaries. This also led to a change in statistical significance. As such, careful consideration around the plausibility of the assumptions required by each estimator is essential.

The choice of estimand should be driven by the trial objectives. We anticipate all four estimands described in Table 2 may be of interest, depending on the specific study objectives. For instance, if interest lies in understanding how well clusters have implemented the intervention (as measured by adherence), a cluster-specific and/or cluster-average estimand may be more appropriate. Conversely, if interest lies in understanding the absolute number of patients that would be saved by the intervention, a marginal participant-average estimand may be most appropriate [4]. We acknowledge that the choice of estimand may not always be straightforward, but this should not discourage conversations between the statisticians and investigators around which estimands may be most appropriate in a given CRT. Further work describing when different estimands would be most appropriate, and using case studies to describe how investigators chose their estimand, would be of value [4].

In this paper we have defined four estimands that could be used in CRTs. However, the estimands described here are not exhaustive; alternative estimands could be defined, for instance by using different weighting schemes than those proposed here. While we feel that the participant- and cluster-average estimands (which give equal weight to participants or clusters) lead to clinically interpretable treatment effects that align well with standard estimators used in CRTs, we acknowledge that other approaches may be of interest to investigators. It is not our intention that investigators must use one of the estimands defined within this paper; in our view, the most important thing is to have a well-defined estimand that is clinically justified based on the study's objectives, and, importantly, this paper provides a coherent framework with standard terminology to enable investigators to describe their estimand, regardless of whether it is one of those described here.

This article suggests a number of areas of future research. For example, we have focused on leveraging the asymptotic properties of each estimator. However, many CRTs only enrol a small number of clusters [23]. Thus, evaluation of the properties of these estimators for well-defined estimands in small sample settings would be useful, for instance by extending previous simulation studies to evaluate these estimators under settings that include informative cluster size [15, 18, 24, 51]). In particular, it may be useful to undertake a comprehensive comparison of the benefits and drawbacks of estimators that require an assumption of 'non-informative cluster size' (i.e. mixed-effects models and GEEs with an exchangeable correlation structure) vs. those that do not (i.e. IEEs and cluster-level summaries). In addition, we have primarily focused on simple parallel-arm CRTs, but longitudinal CRTs with multiple periods are increasingly common. The extension of the estimands in this article to more complex designs, such as the stepped wedge designs [52, 53], may be of interest.

It would also be useful to evaluate the performance of individual methods. For example, evaluation of different variance estimators along with small-sample corrections for the cluster-level analysis approaches would be useful given the model-based variance functions may be misspecified. Additionally, the marginal cluster-level summary estimator we have described has not been extensively studied, and empirical evaluation using simulations (with and without baseline covariates) would be useful.

## 6. Conclusions

Estimands can help clarify research objectives and ensure appropriate statistical estimators are chosen. In cluster-randomised trials, additional attributes of the estimand must be specified compared to individually randomised trials (including whether treatment effects are marginal or cluster-specific, and whether they are participant-average or cluster-average). Choice of these attributes should be based on clinical considerations, and an estimator targeted to the chosen estimand should be used to ensure estimand-aligned statistical analysis of CRTs.

## 7. Appendix

For completeness, we sketch the proof for consistency of each estimator based on the cluster-level summary statistics in Section 3.2. Of note, consistency of the IEE estimators in Section 3.1 is provided in Wang *et al.* [19] (for difference-in-means) and Zhu *et al.* [43] (for odds ratio), and therefore we omit them for brevity. Furthermore, the proof we provide below is based on the super-population perspective as the steps are simpler, relatively more standard in the cluster randomized trials literature, and are more intuitive to understand. The proof under the finite-population perspective requires taking expectation over the randomization distribution and invoking the less familiar finite-population, design-based results [41]. Therefore we only focus on the super-population proof. The consistency results, however, do not change regardless of which of these two perspectives is used.

### 7.1 Consistency of the marginal, participant-average estimator

Recall that we consider a cluster size weighted generalized linear model for the cluster-specific summary statistic $\hat{\pi}_j$ as $logit(\hat{\pi}_j) = \alpha + \beta Z_j$. Below we consider an arbitrary family specification which corresponds to an arbitrary variance function $v(Z_j)$ which is at most a function of treatment; for example, $v(Z_j) \propto 1$ if the family is specified as Gaussian. Under these conditions, the $2 \times 1$ estimating equations can be written out in explicit forms as:

$$\sum_{j=1}^{M} \binom{1}{Z_j} v(Z_j) \left( \frac{\exp(\alpha + \beta Z_j)}{\{1 + \exp(\alpha + \beta Z_j)\}^2} \right) \left( n_j \hat{\pi}_j - n_j \frac{\exp(\alpha + \beta Z_j)}{1 + \exp(\alpha + \beta Z_j)} \right) = 0$$

The first row of this equation is

$$0 = \sum_{j=1}^{M} v(Z_j) \left( \frac{\exp(\alpha + \beta Z_j)}{\{1 + \exp(\alpha + \beta Z_j)\}^2} \right) \left( n_j \hat{\pi}_j - n_j \frac{\exp(\alpha + \beta Z_j)}{1 + \exp(\alpha + \beta Z_j)} \right)$$

$$= \sum_{j=1}^{M} Z_j v(1) \left( \frac{\exp(\alpha + \beta)}{\{1 + \exp(\alpha + \beta)\}^2} \right) \left( n_j \hat{\pi}_j - n_j \frac{\exp(\alpha + \beta)}{1 + \exp(\alpha + \beta)} \right)$$

$$+ \sum_{j=1}^{M} (1 - Z_j) v(0) \left( \frac{\exp(\alpha)}{\{1 + \exp(\alpha)\}^2} \right) \left( n_j \hat{\pi}_j - n_j \frac{\exp(\alpha)}{1 + \exp(\alpha)} \right)$$

Combining with the second row, and since $v(1)$ and $v(0)$ are both constant regardless of the choice of family, they can be omitted and we have:

$$\sum_{j=1}^{M} Z_j v(1) \left( \frac{\exp(\alpha + \beta)}{\{1 + \exp(\alpha + \beta)\}^2} \right) \left( n_j \hat{\pi}_j - n_j \frac{\exp(\alpha + \beta)}{1 + \exp(\alpha + \beta)} \right)$$

$$= \sum_{j=1}^{M} Z_j \left( n_j \hat{\pi}_j - n_j \frac{\exp(\alpha + \beta)}{1 + \exp(\alpha + \beta)} \right) = 0$$

$$\sum_{j=1}^{M} (1 - Z_j) v(0) \left( \frac{\exp(\alpha)}{\{1 + \exp(\alpha)\}^2} \right) \left( n_j \hat{\pi}_j - n_j \frac{\exp(\alpha)}{1 + \exp(\alpha)} \right)$$

$$= \sum_{j=1}^{M} (1 - Z_j) \left( n_j \hat{\pi}_j - n_j \frac{\exp(\alpha)}{1 + \exp(\alpha)} \right) = 0$$

Solving these two equations for the parameters and rearranging terms, we obtain the form of the estimator as:

$$\hat{\beta} = logit \left\{ \frac{\sum_{j=1}^{M} Z_j n_j \hat{\pi}_j}{\sum_{j=1}^{M} Z_j n_j} \right\} - logit \left\{ \frac{\sum_{j=1}^{M} (1 - Z_j) n_j \hat{\pi}_j}{\sum_{j=1}^{M} (1 - Z_j) n_j} \right\} = logit\{\hat{P}_1\} - logit\{\hat{P}_0\}$$

Now observe that, as the number of clusters increases to infinity ($M \to \infty$)

$$\hat{P}_1 = \frac{M^{-1} \sum_{j=1}^{M} \sum_{i=1}^{n_j} Z_j Y_{ij}}{M^{-1} \sum_{j=1}^{M} Z_j n_j} \xrightarrow{p} \frac{E\left[\sum_{i=1}^{n_j} Z_j Y_{ij}\right]}{E[Z_j n_j]} = \frac{E(Z_j) E\left[\sum_{i=1}^{n_j} Y_{ij}^{(1)}\right]}{E(Z_j) E[n_j]} = \frac{E\left[\sum_{i=1}^{n_j} Y_{ij}^{(1)}\right]}{E[n_j]}$$

where the convergence in probability statement results from an application of the Weak Law of Large Numbers for independent but non-identically distributed data and the Continuous Mapping Theorem. The subsequent equality is due to cluster-level randomization. Following the exact same steps under control, we observe that:

$$\hat{P}_0 \xrightarrow{p} \frac{E\left[\sum_{i=1}^{n_j} Y_{ij}^{(0)}\right]}{E[n_j]}$$

and therefore, by the Continuous Mapping Theorem:

$$\exp(\hat{\beta}) \xrightarrow{p} \frac{\dfrac{E(\sum_{i=1}^{n_j} Y_{ij}^{(1)})}{E(n_j)} \Big/ \left(1 - \dfrac{E(\sum_{i=1}^{n_j} Y_{ij}^{(1)})}{E(n_j)}\right)}{\dfrac{E(\sum_{i=1}^{n_j} Y_{ij}^{(0)})}{E(n_j)} \Big/ \left(1 - \dfrac{E(\sum_{i=1}^{n_j} Y_{ij}^{(0)})}{E(n_j)}\right)} = \Gamma^{MG-PA}.$$

Clearly, since $v(Z_j)$ does not enter into the estimating equations and hence the treatment effect estimator, the choice of family has no impact on the final treatment effect estimator.

7.2 Consistency of the marginal, cluster-average estimator

To find the form of the estimator of the unweighted cluster-level generalized linear model, we simply remove the cluster size weight in the estimating equation of Section 7.1, and obtain:

$$\sum_{j=1}^{M} \binom{1}{Z_j} v(Z_j) \left(\frac{\exp(\alpha + \beta Z_j)}{\{1 + \exp(\alpha + \beta Z_j)\}^2}\right) \left(\hat{\pi}_j - \frac{\exp(\alpha + \beta Z_j)}{1 + \exp(\alpha + \beta Z_j)}\right) = 0$$

Since $v(Z_j)$ is only at most a function of the treatment indicator, the above equation then implies:

$$\sum_{j=1}^{M} Z_j \left(\hat{\pi}_j - \frac{\exp(\alpha + \beta)}{1 + \exp(\alpha + \beta)}\right) = 0$$

$$\sum_{j=1}^{M} (1 - Z_j) \left(\hat{\pi}_j - \frac{\exp(\alpha)}{1 + \exp(\alpha)}\right) = 0$$

and the treatment effect estimator is given explicitly as:

$$\hat{\beta} = \text{logit}\left\{\frac{\sum_{j=1}^{M} Z_j \hat{\pi}_j}{\sum_{j=1}^{M} Z_j}\right\} - \text{logit}\left\{\frac{\sum_{j=1}^{M}(1 - Z_j)\hat{\pi}_j}{\sum_{j=1}^{M}(1 - Z_j)}\right\}$$

$$= \text{logit}\{\hat{P}_1\} - \text{logit}\{\hat{P}_0\} \xrightarrow{p} \text{logit}\left\{E\left(\frac{\sum_{i=1}^{n_j} Y_{ij}^{(1)}}{n_j}\right)\right\} - \text{logit}\left\{E\left(\frac{\sum_{i=1}^{n_j} Y_{ij}^{(0)}}{n_j}\right)\right\}$$

where the convergence in probability statements assumes the number of clusters $M \to \infty$, and is a result of appropriate Weak Law of Large Numbers and the Continuous Mapping Theorem. Therefore:

$$\exp(\hat{\beta}) \xrightarrow{p} \frac{E\left(\dfrac{\sum_{i=1}^{n_j} Y_{ij}^{(1)}}{n_j}\right) \Big/ \left(1 - E\left(\dfrac{\sum_{i=1}^{n_j} Y_{ij}^{(1)}}{n_j}\right)\right)}{E\left(\dfrac{\sum_{i=1}^{n_j} Y_{ij}^{(0)}}{n_j}\right) \Big/ \left(1 - E\left(\dfrac{\sum_{i=1}^{n_j} Y_{ij}^{(0)}}{n_j}\right)\right)} = \Gamma^{MG-CA}.$$

Similarly, the choice of family has no impact on the final treatment effect estimator.

7.3 Consistency of the cluster-specific, participant-average estimator

Recall that a cluster size weighted linear regression is fitted to the log odds ratio summary in each cluster, such that a working model is given by $log(\hat{\pi}_j/\{1-\hat{\pi}_j\}) = \alpha + \beta Z_j + \varepsilon_j$ in Section 3.2.3. Denote the $M \times 2$ design matrix as $D = (1_M, Z)$, where $1_M$ is a $M$-vector and $Z = (Z_1, \ldots Z_M)'$ is the vector of treatment indicators; define $\hat{Y}^{smr} = (\hat{Y}_1^{smr}, \ldots, \hat{Y}_M^{smr})' = (log(\hat{\pi}_1/\{1-\hat{\pi}_1\}), \ldots log(\hat{\pi}_M/\{1-\hat{\pi}_M\}))'$ as the vector of responses, and $W = diag\{n_j\}$ as the weight matrix. Then we have:

$$(D'WD)^{-1} = \begin{pmatrix} N & \sum_{j=1}^{M} Z_j n_j \\ \sum_{j=1}^{M} Z_j n_j & \sum_{j=1}^{M} Z_j n_j \end{pmatrix}^{-1}$$

$$= \frac{1}{N \sum_{j=1}^{M} Z_j n_j - (\sum_{j=1}^{M} Z_j n_j)^2} \begin{pmatrix} \sum_{j=1}^{M} Z_j n_j & -\sum_{j=1}^{M} Z_j n_j \\ -\sum_{j=1}^{M} Z_j n_j & N \end{pmatrix},$$

$$D'W\hat{Y}^{smr} = \begin{pmatrix} \sum_{j=1}^{M} \hat{Y}_j^{smr} n_j \\ \sum_{j=1}^{M} Z_j \hat{Y}_j^{smr} n_j \end{pmatrix}.$$

Therefore, the treatment coefficient estimator $\hat{\beta}$ is the last element of the weighted least squares formula, given by:

$$\hat{\beta} = \{(D'WD)^{-1}(D'W\hat{Y}^{smr})\}_{[2,1]} = \frac{N \sum_{j=1}^{M} Z_j \hat{Y}_j^{smr} n_j - (\sum_{j=1}^{M} \hat{Y}_j^{smr} n_j)(\sum_{j=1}^{M} Z_j n_j)}{N \sum_{j=1}^{M} Z_j n_j - (\sum_{j=1}^{M} Z_j n_j)^2}$$

$$= \frac{(\sum_{j=1}^{M}(1-Z_j)n_j)(\sum_{j=1}^{M} Z_j \hat{Y}_j^{smr} n_j) - (\sum_{j=1}^{M}(1-Z_j)\hat{Y}_j^{smr} n_j)(\sum_{j=1}^{M} Z_j n_j)}{(\sum_{j=1}^{M} Z_j n_j)(\sum_{j=1}^{M}(1-Z_j)n_j)}$$

$$= \frac{\sum_{j=1}^{M} Z_j \hat{Y}_j^{smr} n_j}{\sum_{j=1}^{M} Z_j n_j} - \frac{\sum_{j=1}^{M}(1-Z_j)\hat{Y}_j^{smr} n_j}{\sum_{j=1}^{M}(1-Z_j)n_j} \quad (20)$$

Define:

$$ODDS_j(z) = \frac{1}{n_j}\sum_{i=1}^{n_j} Y_{ij}^{(z)} / \left(1 - \frac{1}{n_j}\sum_{i=1}^{n_j} Y_{ij}^{(z)}\right)$$

and therefore $OR_j = \frac{ODDS_j(1)}{ODDS_j(0)}$. Under the consistency assumption,

$\hat{Y}_j^{smr} = \log\left(\frac{\hat{\pi}_j}{1-\hat{\pi}_j}\right) = Z_j \log(ODDS_j(1)) + (1-Z_j)\log\left(ODDS_j(0)\right)$, and therefore (20) becomes

$$\hat{\beta} = \frac{\sum_{j=1}^{M} Z_j \log(ODDS_j(1)) n_j}{\sum_{j=1}^{M} Z_j n_j} - \frac{\sum_{j=1}^{M}(1-Z_j) \log\left(ODDS_j(0)\right) n_j}{\sum_{j=1}^{M}(1-Z_j) n_j} \xrightarrow{p} \frac{E\left[Z_j \log\left(ODDS_j(1)\right) n_j\right]}{E[Z_j n_j]}$$

$$-\frac{E\left[(1-Z_j)\log\left(ODDS_j(0)\right) n_j\right]}{E[(1-Z_j) n_j]} = \frac{E\left[n_j \left\{\log\left(ODDS_j(1)/ODDS_j(0)\right)\right\}\right]}{E[n_j]}$$

$$= \frac{E[n_j \{\log(OR_j)\}]}{E[n_j]} = \log(\Gamma^{CS-PA}).$$

In the above, the convergence in probability statement results from an application of the Weak Law of Large Numbers for independent but non-identically distributed data, the subsequent equality is due to randomization of $Z_j$ such that $E\left[Z_j \log\left(ODDS_j(1)\right) n_j\right] = E(Z_j) E\left[\log\left(ODDS_j(1)\right) n_j\right]$ and $E[Z_j n_j] = E(Z_j) E[n_j]$ etc. Then by the Continuous Mapping Theorem, we have $\exp(\hat{\beta}) \xrightarrow{p} \Gamma^{CS-PA}$.

7.4 Consistency of the cluster-specific, cluster-average estimator

The proof of consistency follows from Section 7.3 by replacing $n_j = 1$ for all j. Therefore, we obtain that:

$$\hat{\beta} = \frac{\sum_{j=1}^{M} Z_j \log(ODDS_j(1))}{\sum_{j=1}^{M} Z_j} - \frac{\sum_{j=1}^{M}(1-Z_j) \log\left(ODDS_j(0)\right)}{\sum_{j=1}^{M}(1-Z_j)} \xrightarrow{p} E[\log(OR_j)] = \log(\Gamma^{CS-CA}).$$

This then leads to $\exp(\hat{\beta}) \xrightarrow{p} \Gamma^{CS-C}$ under the Continuous Mapping Theorem.

**Figure 1 – Difference between marginal participant- vs. cluster-average odds ratio for "postextubation stridor" in the Randomized Evaluation of Sedation Titration for Respiratory Failure (RESTORE) trial.** Each bubble denotes the proportion of events in that cluster. The size of the bubbles represent the weight given to each cluster, with red bubbles representing the participant-average weighting, and blue bubbles denoting the cluster-average weighting. The overall treatment group means are closer together under the cluster-average weighting than the participant-average weighting, owing to the cluster-average weighting giving more weight to smaller clusters with smaller between-group differences.

**Table 1 – Difference between marginal/cluster-specific and participant-/cluster-average attributes of estimand**

| Concept | Description |
|---|---|
| **Marginal vs. cluster-specific estimands** | |
| *Marginal* | A *marginal* estimand (also called "population-averaged") is where the potential outcomes are first summarised separately by treatment condition, and then these summaries are contrasted between treatment conditions to obtain an overall treatment effect. |
| *Cluster-specific* | A *cluster-specific* estimand (also called "conditional") is where the potential outcomes are summarised and contrasted within each cluster to obtain cluster-specific treatment effects, and then an average of these cluster-specific effects is calculated to obtain an overall treatment effect (note this average could be taken in different ways; see participant- vs. cluster-average estimands). |
| **Participant- vs. cluster-average estimands** | |
| *Participant-average* | A *participant-average* estimand is where each participant is given equal weight |
| *Cluster-average* | A *cluster-average* estimand is where each cluster is given equal weight |

**Table 2 – Overview of estimands for cluster-randomised trials[a]**

| | Super-population perspective[b] | | Finite-population perspective[b] | |
|---|---|---|---|---|
| Estimand[c] | Definition for a difference[d] | Definition for an odds ratio | Definition for a difference[d] | Definition for an odds ratio |
| Marginal, participant-average ($\Delta^{MG-}$ and $\Gamma^{MG-}$) | $= \frac{E(\sum_{i=1}^{n_j} Y_{ij}^{(1)})}{E(n_j)} - \frac{E(\sum_{i=1}^{n_j} Y_{ij}^{(0)})}{E(n_j)}$ | $\frac{E(\sum_{i=1}^{n_j} Y_{ij}^{(1)})}{E(n_j)} / \left(1 - \frac{E(\sum_{i=1}^{n_j} Y_{ij}^{(1)})}{E(n_j)}\right)$ $\overline{\frac{E(\sum_{i=1}^{n_j} Y_{ij}^{(0)})}{E(n_j)} / \left(1 - \frac{E(\sum_{i=1}^{n_j} Y_{ij}^{(0)})}{E(n_j)}\right)}$ | $\frac{1}{N} \sum_{j=1}^{M} \sum_{i=1}^{n_j} Y_{ij}^{(1)} - \frac{1}{N} \sum_{j=1}^{M} \sum_{i=1}^{n_j} Y_{ij}^{(0)}$ | $\frac{\frac{1}{N} \sum_{j=1}^{M} \sum_{i=1}^{n_j} Y_{ij}^{(1)} / \left(1 - \frac{1}{N} \sum_{j=1}^{M} \sum_{i=1}^{n_j} Y_{ij}^{(1)}\right)}{\frac{1}{N} \sum_{j=1}^{M} \sum_{i=1}^{n_j} Y_{ij}^{(0)} / \left(1 - \frac{1}{N} \sum_{j=1}^{M} \sum_{i=1}^{n_j} Y_{ij}^{(0)}\right)}$ |
| Marginal, cluster-average ($\Delta^{MG-CA}$ and $\Gamma^{MG-CA}$) | $E\left(\frac{\sum_{i=1}^{n_j} Y_{ij}^{(1)}}{n_j}\right) - E\left(\frac{\sum_{i=1}^{n_j} Y_{ij}^{(0)}}{n_j}\right)$ | $\frac{E\left(\frac{\sum_{i=1}^{n_j} Y_{ij}^{(1)}}{n_j}\right) / \left(1 - E\left(\frac{\sum_{i=1}^{n_j} Y_{ij}^{(1)}}{n_j}\right)\right)}{E\left(\frac{\sum_{i=1}^{n_j} Y_{ij}^{(0)}}{n_j}\right) / \left(1 - E\left(\frac{\sum_{i=1}^{n_j} Y_{ij}^{(0)}}{n_j}\right)\right)}$ | $\frac{1}{M} \sum_{j=1}^{M} \bar{Y}_j^{(1)} - \frac{1}{M} \sum_{j=1}^{M} \bar{Y}_j^{(0)}$ | $\frac{\frac{1}{M} \sum_{j=1}^{M} \bar{Y}_j^{(1)} / \left(1 - \frac{1}{M} \sum_{j=1}^{M} \bar{Y}_j^{(1)}\right)}{\frac{1}{M} \sum_{j=1}^{M} \bar{Y}_j^{(0)} / \left(1 - \frac{1}{M} \sum_{j=1}^{M} \bar{Y}_j^{(0)}\right)}$ |
| Cluster-specific, participant-average[e] ($\Delta^{CS-PA}$ and $\Gamma^{CS-PA}$) | $\frac{E(n_j \beta_j)}{E(n_j)}$ <br> Where: <br> $\beta_j = \bar{Y}_j^{(1)} - \bar{Y}_j^{(0)}$ | $\frac{E(n_j f(OR_j))}{E(n_j)}$ <br> Where: <br> $OR_j = \frac{\frac{1}{n_j} \sum_{i=1}^{n_j} Y_{ij}^{(1)} / \left(1 - \frac{1}{n_j} \sum_{i=1}^{n_j} Y_{ij}^{(1)}\right)}{\frac{1}{n_j} \sum_{i=1}^{n_j} Y_{ij}^{(0)} / \left(1 - \frac{1}{n_j} \sum_{i=1}^{n_j} Y_{ij}^{(0)}\right)}$ | $\frac{\sum_{j=1}^{M} n_j \beta_j}{\sum_{j=1}^{M} n_j}$ <br> Where: <br> $\beta_j = \bar{Y}_j^{(1)} - \bar{Y}_j^{(0)}$ | $\frac{\sum_{j=1}^{M} n_j f(OR_j)}{\sum_{j=1}^{M} n_j}$ <br> Where: <br> $OR_j = \frac{\frac{1}{n_j} \sum_{i=1}^{n_j} Y_{ij}^{(1)} / \left(1 - \frac{1}{n_j} \sum_{i=1}^{n_j} Y_{ij}^{(1)}\right)}{\frac{1}{n_j} \sum_{i=1}^{n_j} Y_{ij}^{(0)} / \left(1 - \frac{1}{n_j} \sum_{i=1}^{n_j} Y_{ij}^{(0)}\right)}$ |
| Cluster-specific, cluster-average[e] ($\Delta^{CS-CA}$ | $E(\beta_j)$ <br> Where: <br> $\beta_j = \bar{Y}_j^{(1)} - \bar{Y}_j^{(0)}$ | $E(f(OR_j))$ <br> Where: | $\frac{1}{M} \sum_{j=1}^{M} \beta_j$ <br> Where: | $\frac{1}{M} \sum_{j=1}^{M} f(OR_j)$ <br> Where: |
| and $\Gamma^{CS-CA}$) | | $OR_j = \dfrac{\frac{1}{n_j}\sum_{i=1}^{n_j} Y_{ij}^{(1)} / \left(1 - \frac{1}{n_j}\sum_{i=1}^{n_j} Y_{ij}^{(1)}\right)}{\frac{1}{n_j}\sum_{i=1}^{n_j} Y_{ij}^{(0)} / \left(1 - \frac{1}{n_j}\sum_{i=1}^{n_j} Y_{ij}^{(0)}\right)}$ | $\beta_j = \bar{Y}_j^{(1)} - \bar{Y}_j^{(0)}$ | $OR_j = \dfrac{\frac{1}{n_j}\sum_{i=1}^{n_j} Y_{ij}^{(1)} / \left(1 - \frac{1}{n_j}\sum_{i=1}^{n_j} Y_{ij}^{(1)}\right)}{\frac{1}{n_j}\sum_{i=1}^{n_j} Y_{ij}^{(0)} / \left(1 - \frac{1}{n_j}\sum_{i=1}^{n_j} Y_{ij}^{(0)}\right)}$ |

[a] $M$ and $N$ represent the total number of clusters and the total number of participants respectively, $Y_{ij}^{(1)}$ and $Y_{ij}^{(0)}$ denote potential outcomes from participant $i$ in cluster $j$ under intervention and control respectively, $n_j$ denotes the size of cluster $j$, and $\bar{Y}_j^{(1)} = \frac{1}{n_j}\sum_{i=1}^{n_j} Y_{ij}^{(1)}$ (and similarly for $\bar{Y}_j^{(0)}$).

[b] Under the super-population perspective, each cluster is assumed to be an independent random sample from a hypothetical infinite population of clusters, and the expectation for each estimand is defined with respect to that super-population of clusters. Under the finite-population perspective, the trial sample itself is considered as the finite target population, and so the estimand is defined for these clusters. Mathematically, the difference between the two frameworks is that one uses population expectation for the estimand while the other considers a finite-sample estimand.

[c] In order to be fully defined, estimands require further specification of the (i) population; (ii) treatment conditions; (iii) endpoint; (iv) summary measure; and (v) handling of intercurrent events

[d] For differences, marginal and cluster-specific estimands coincide

[e] $f(OR_j)$ is some function of the cluster-specific odds ratio, for instance identity or log. Each function corresponds to a differ way of averaging the cluster-specific odds ratios across clusters. For instance, an identity function, where $f(OR_j) = OR_j$ would just take a simple average of the odds ratios. A log function, where $f(OR_j) = log(OR_j)$ would take the average of the log-odds ratios (importantly, the resulting average would need to be back-transformed to the odds ratio scale, e.g. by taking $exp\left(\frac{\sum_{j=1}^{M} n_j f(OR_j)}{\sum_{j=1}^{M} n_j}\right)$ or $exp\left(\frac{1}{M}\sum_{j=1}^{M} f(OR_j)\right)$. We recommend using $f(OR_j) = log(OR_j)$ as this typically leads to more interpretable results; see Section 2.4.3 and Section 2.4.4.

**Table 3 – Overview of estimators for each estimand**

| Estimand | Description | Estimator | Description | Assumptions |
|---|---|---|---|---|
| Marginal, participant-average | Average effect across participants<br><br>Potential outcomes are summarised under each treatment condition, and these summaries are contrasted | IEEs (unweighted) | Individual participant outcomes are analysed using an independent working correlation structure in conjunction with cluster-robust standard errors.<br><br>Can be implemented using (a) GEEs with a working independence correlation structure; or (b) using maximum likelihood/least squares estimators (both options, (a) and (b), using cluster-robust standard errors); importantly, the models are unweighted. | Standard assumptions[a] |
| | | Cluster-level summaries (weighted) | A summary measure is taken in each cluster (e.g. the proportion), and an appropriate weighted generalised linear model is used for analysis, with the cluster-level summaries as the outcomes, and weights equal to the cluster-size $n_j$.[b]<br><br>For instance, a logit link could be used for the odds ratio; a family should be chosen such that the variance is independent of the mean (e.g. Gaussian). | Standard assumptions[a] |
| | | GEEs with exchangeable correlation (unweighted) | GEEs are applied to individual participant data, using an exchangeable working correlation structure, in conjunction with cluster-robust standard errors. | Standard assumptions[a]<br><br>Non-informative cluster size |
| Marginal, cluster-average | Average effect across clusters<br><br>Potential outcomes are summarised under each treatment condition, and these summaries are contrasted | IEEs (weighted) | As described above (under "*Marginal, participant average*" estimand); however, participants are weighted by their inverse cluster-size, $\frac{1}{n_j}$.[b] | Standard assumptions[a] |

| | | Cluster-level summaries (unweighted) | As described above (under "*Marginal, participant average*" estimand); however, the generalised linear model is unweighted. | Standard assumptions[a] |
|---|---|---|---|---|
| Cluster-specific, participant-average | Average effect across participants<br><br>Potential outcomes are summarised and contrasted within each cluster separately, and these cluster-specific effects are then averaged, giving equal weight to each participant | Cluster-level summaries (weighted) | First, a summary measure is taken in each cluster (e.g. a proportion). Next, the summary measure is transformed if necessary; for instance, if the aim is to estimate an odds ratio, the log-odds is taken for each cluster. Finally, a weighted linear regression model is applied to the cluster-specific (transformed) summary measures, with weights equal to the cluster-size $n_j$[b].<br><br>The estimate may then need to be back-transformed (e.g. if the log-odds have been used, the odds ratio is calculated by exponentiating the estimate). | Standard assumptions[a]<br><br>For binary outcomes, the average outcome in each cluster is bounded away from 0 or 1 |
| | | Mixed-effects model | A mixed-effects model is applied to individual participant data, with a random-intercept for cluster. | Standard assumptions[a]<br><br>Non-informative cluster size |
| Cluster-specific, cluster-average | Average effect across clusters<br><br>Potential outcomes are summarised and contrasted within each cluster separately, and these cluster-specific effects are then averaged, giving equal weight to each cluster | Cluster-level summaries (unweighted) | As described above (under "*Cluster-specific, participant average*" estimand); however, the regression model is unweighted. | Standard assumptions[a]<br><br>For binary outcomes, the average outcome in each cluster is bounded away from 0 or 1 |

[a] Standard assumptions are: (a) the consistency assumption (sometimes termed the stable unit treatment value assumption), which requires that $Y_{ij} = Y_{ij}^{(z)}$ for $Z = z$, i.e. that a participant's observed outcome is equal to their potential outcome under their allocated treatment; (b) the maximum cluster size is bounded; (c) exchangeability between treatment arms (i.e. that clusters are randomised between treatment arms, and there is no differential enrolment

of patients between treatments); (d) observations are independent across clusters; and (e) a large number of clusters such that appropriate versions of Law of Large Numbers and Central Limit Theorems can be applied.

[b] The cluster-size $n_j$ is based on the number of participants in each cluster included in the analysis.

**Table 4 – Re-analysis results for the Randomized Evaluation of Sedation Titration for Respiratory Failure (RESTORE) trial for the outcome "postextubation stridor".** Odds ratios are for the comparison intervention vs. control. This trial enrolled 31 clusters, and the event rate was 7.2% in the intervention vs. 4.5% in the control. ICC=0.03. P-value for interaction between treatment and cluster size ≥100 = 0.21. Marginal, participant-average odds ratio = 1.15 (0.56 to 2.36) in clusters of size <100, and 2.26 (1.49 to 3.43) in clusters of size ≥100.

| Estimand | Estimator | Odds ratio (95% CI) | P-value |
|---|---|---|---|
| Marginal, participant-average | | | |
| | IEEs (unweighted)[a] | 1.65 (1.02 to 2.67) | 0.04 |
| | Cluster-level summaries (weighted)[b] | 1.65 (1.08 to 2.51) | 0.02 |
| | GEEs with exchangeable correlation (unweighted)[a] | 1.57 (0.98 to 2.50) | 0.06 |
| Cluster-specific, participant-average | | | |
| | Cluster-level summaries (weighted)[b] | 1.83 (1.06 to 3.14) | 0.03 |
| | Mixed-effects logistic regression model | 1.54 (0.97 to 2.44) | 0.07 |
| Marginal, cluster-average | | | |
| | IEEs (weighted)[a] | 1.38 (0.87 to 2.19) | 0.17 |
| | Cluster-level summaries (unweighted)[b] | 1.38 (0.89 to 2.14) | 0.15 |
| Cluster-specific, cluster-average | | | |
| | Cluster-level summaries (unweighted)[b] | 1.51 (0.90 to 2.52) | 0.11 |

[a] Confidence intervals were calculated using cluster-robust standard errors with the Fay-Graubard correction

[b] Confidence intervals were calculated using Huber-White standard errors

**References**

1. ICH E9 (R1) addendum on estimands and sensitivity analysis in clinical trials to the guideline on statistical principles for clinical trials [Available from: https://www.ema.europa.eu/en/documents/scientific-guideline/ich-e9-r1-addendum-estimands-sensitivity-analysis-clinical-trials-guideline-statistical-principles_en.pdf.
2. Kahan BC, Cro S, Li F, Harhay MO. Eliminating Ambiguous Treatment Effects Using Estimands. American Journal of Epidemiology. 2023;192(6):987-94.
3. Kahan BC, Hindley J, Edwards M, Cro S, Morris TP. The estimands framework: a primer on the ICH E9(R1) addendum. BMJ. 2024;384:e076316.
4. Kahan BC, Li F, Copas AJ, Harhay MO. Estimands in cluster-randomized trials: choosing analyses that answer the right question. International Journal of Epidemiology. 2022.
5. Kahan BC, Morris TP, White IR, Carpenter J, Cro S. Estimands in published protocols of randomised trials: urgent improvement needed. Trials. 2021;22(1):686.
6. Roydhouse J, Floden L, Braat S, Grobler A, Kochovska S, Currow DC, Bell ML. Missing data in palliative care research: estimands and estimators. BMJ Supportive & Palliative Care. 2022;12(4):464-70.
7. Keene ON, Wright D, Phillips A, Wright M. Why ITT analysis is not always the answer for estimating treatment effects in clinical trials. Contemporary Clinical Trials. 2021;108:106494.
8. Cro S, Kahan BC, Rehal S, Chis Ster A, Carpenter JR, White IR, Cornelius VR. Evaluating how clear the questions being investigated in randomised trials are: systematic review of estimands. BMJ. 2022;378:e070146.
9. Mitroiu M, Teerenstra S, Oude Rengerink K, Pétavy F, Roes KCB. Estimation of treatment effects in short-term depression studies. An evaluation based on the ICH E9(R1) estimands framework. Pharmaceutical Statistics.n/a(n/a).
10. Lynggaard H, Bell J, Lösch C, Besseghir A, Rantell K, Schoder V, Lanius V. Principles and recommendations for incorporating estimands into clinical study protocol templates. Trials. 2022;23(1):685.
11. Fletcher C, Hefting N, Wright M, Bell J, Anzures-Cabrera J, Wright D, et al. Marking 2-Years of New Thinking in Clinical Trials: The Estimand Journey. Therapeutic Innovation & Regulatory Science. 2022;56(4):637-50.
12. Campbell MK, Piaggio G, Elbourne DR, Altman DG, Group C. Consort 2010 statement: extension to cluster randomised trials. BMJ. 2012;345:e5661.
13. Donner A, Klar N. Design and Analysis of Cluster Randomization Trials in Health Research. London: Arnold; 2000.
14. Eldridge S, Kerry S. A practical guide to cluster randomised trials in health services research. Chichester: Wiley; 2012.
15. Hayes RJ, Moulton LH. Cluster Randomised Trials: Chapman and Hall/CRC; 2017.
16. Murray DM. Design and Analysis of Group-Randomized Trials. Monographs in Epidemiology and iostatistics. Oxford: Oxford University Press; 1998.
17. Turner EL, Li F, Gallis JA, Prague M, Murray DM. Review of Recent Methodological Developments in Group-Randomized Trials: Part 1-Design. Am J Public Health. 2017;107(6):907-15.
18. Turner EL, Prague M, Gallis JA, Li F, Murray DM. Review of Recent Methodological Developments in Group-Randomized Trials: Part 2-Analysis. Am J Public Health. 2017;107(7):1078-86.
19. Wang X, Turner EL, Li F, Wang R, Moyer J, Cook AJ, et al. Two weights make a wrong: Cluster randomized trials with variable cluster sizes and heterogeneous treatment effects. Contemp Clin Trials. 2022;114:106702.
20. Ma J, Raina P, Beyene J, Thabane L. Comparison of population-averaged and cluster-specific models for the analysis of cluster randomized trials with missing binary outcomes: a simulation study. BMC Medical Research Methodology. 2013;13(1):9.

21. Ukoumunne OC, Carlin JB, Gulliford MC. A simulation study of odds ratio estimation for binary outcomes from cluster randomized trials. Statistics in medicine. 2007;26(18):3415-28.
22. Turner EL, Platt AC, Gallis JA, Tetreault K, Easter C, McKenzie JE, et al. Completeness of reporting and risks of overstating impact in cluster randomised trials: a systematic review. The Lancet Global Health. 2021;9(8):e1163-e8.
23. Kahan BC, Forbes G, Ali Y, Jairath V, Bremner S, Harhay MO, et al. Increased risk of type I errors in cluster randomised trials with small or medium numbers of clusters: a review, reanalysis, and simulation study. Trials. 2016;17(1):438.
24. Leyrat C, Morgan KE, Leurent B, Kahan BC. Cluster randomized trials with a small number of clusters: which analyses should be used? Int J Epidemiol. 2018;47(1):321-31.
25. Benitez A, Petersen ML, van der Laan MJ, Santos N, Butrick E, Walker D, et al. Defining and Estimating Effects in Cluster Randomized Trials: A Methods Comparison2021 October 01, 2021:[arXiv:2110.09633 p.]. Available from: https://ui.adsabs.harvard.edu/abs/2021arXiv211009633B.
26. Bugni F, Canay I, Shaikh A, Tabord-Meehan M. Inference for Cluster Randomized Experiments with Non-ignorable Cluster Sizes2022 April 01, 2022:[arXiv:2204.08356 p.]. Available from: https://ui.adsabs.harvard.edu/abs/2022arXiv220408356B.
27. Kosuke I, Gary K, Clayton N. The Essential Role of Pair Matching in Cluster-Randomized Experiments, with Application to the Mexican Universal Health Insurance Evaluation. Statistical Science. 2009;24(1):29-53.
28. Su F, Ding P. Model-Assisted Analyses of Cluster-Randomized Experiments. Journal of the Royal Statistical Society Series B: Statistical Methodology. 2021;83(5):994-1015.
29. Wang B, Park C, Small DS, Li F. Model-Robust and Efficient Covariate Adjustment for Cluster-Randomized Experiments. Journal of the American Statistical Association.1-13.
30. Clark TP, Kahan BC, Phillips A, White I, Carpenter JR. Estimands: bringing clarity and focus to research questions in clinical trials. BMJ Open. 2022;12(1):e052953.
31. Li F, Tian Z, Bobb J, Papadogeorgou G, Li F. Clarifying selection bias in cluster randomized trials. Clinical Trials. 2021:17407745211056875.
32. Li F, Tian Z, Tian Z, Li F. A note on identification of causal effects in cluster randomized trials with post-randomization selection bias. Communications in Statistics - Theory and Methods. 2022:1-13.
33. Kenny A, Voldal EC, Xia F, Heagerty PJ, Hughes JP. Analysis of stepped wedge cluster randomized trials in the presence of a time-varying treatment effect. Statistics in medicine. 2022;41(22):4311-39.
34. Maleyeff L, Li F, Haneuse S, Wang R. Assessing exposure-time treatment effect heterogeneity in stepped-wedge cluster randomized trials. Biometrics.n/a(n/a).
35. Bai Y, Liu J, Shaikh AM, Tabord-Meehan M. Inference in Cluster Randomized Trials with Matched Pairs2022 November 01, 2022:[arXiv:2211.14903 p.]. Available from: https://ui.adsabs.harvard.edu/abs/2022arXiv221114903B.
36. Huitfeldt A, Stensrud MJ, Suzuki E. On the collapsibility of measures of effect in the counterfactual causal framework. Emerg Themes Epidemiol. 2019;16:1.
37. Seaman S, Pavlou M, Copas A. Review of methods for handling confounding by cluster and informative cluster size in clustered data. Statistics in medicine. 2014;33(30):5371-87.
38. Seaman SR, Pavlou M, Copas AJ. Methods for observed-cluster inference when cluster size is informative: A review and clarifications. Biometrics. 2014;70(2):449-56.
39. Balzer LB, van der Laan M, Ayieko J, Kamya M, Chamie G, Schwab J, et al. Two-Stage TMLE to reduce bias and improve efficiency in cluster randomized trials. Biostatistics. 2023;24(2):502-17.
40. Balzer LB, Zheng W, van der Laan MJ, Petersen ML. A new approach to hierarchical data analysis: Targeted maximum likelihood estimation for the causal effect of a cluster-level exposure. Stat Methods Med Res. 2019;28(6):1761-80.

41. Li X, Ding P. General Forms of Finite Population Central Limit Theorems with Applications to Causal Inference. Journal of the American Statistical Association. 2017;112(520):1759-69.
42. Vaart AWvd. Asymptotic Statistics. Cambridge: Cambridge University Press; 1998.
43. Zhu AY, Mitra N, Hemming K, Harhay MO, Li F. Leveraging baseline covariates to analyze small cluster-randomized trials with a rare binary outcome. Biom J. 2023:e2200135.
44. Hernán MA, Robins JM. Causal Inference: What If. Boca Raton: Chapman & Hall/CRC; 2020.
45. Liang K-Y, Zeger SL. Longitudinal data analysis using generalized linear models. Biometrika. 1986;73(1):13-22.
46. Wooldridge JM. Econometric Analysis of Cross Section and Panel Data: The MIT Press; 2010.
47. Wang B, Harhay MO, Small DS, Morris TP, Li F. On the mixed-model analysis of covariance in cluster-randomized trials2021 December 01, 2021:[arXiv:2112.00832 p.]. Available from: https://ui.adsabs.harvard.edu/abs/2021arXiv211200832W.
48. Curley MA, Wypij D, Watson RS, Grant MJ, Asaro LA, Cheifetz IM, et al. Protocolized sedation vs usual care in pediatric patients mechanically ventilated for acute respiratory failure: a randomized clinical trial. JAMA. 2015;313(4):379-89.
49. Fay MP, Graubard BI. Small-sample adjustments for Wald-type tests using sandwich estimators. Biometrics. 2001;57(4):1198-206.
50. Kahan BC, Li F, Blette B, Jairath V, Copas A, Harhay M. Informative cluster size in cluster-randomised trials: A case study from the TRIGGER trial. Clinical trials (London, England). 2023:17407745231186094.
51. Thompson JA, Leyrat C, Fielding KL, Hayes RJ. Cluster randomised trials with a binary outcome and a small number of clusters: comparison of individual and cluster level analysis method. BMC Medical Research Methodology. 2022;22(1):222.
52. Chen X, Li F. Model-assisted analysis of covariance estimators for stepped wedge cluster randomized experiments2023 June 01, 2023:[arXiv:2306.11267 p.]. Available from: https://ui.adsabs.harvard.edu/abs/2023arXiv230611267C.
53. Wang B, Wang X, Wang R, Li F. How to achieve model-robust inference in stepped wedge trials with model-based methods?2024 January 01, 2024:[arXiv:2401.15680 p.]. Available from: https://ui.adsabs.harvard.edu/abs/2024arXiv240115680W.

**Funding** B.C.K. and A.J.C. are funded by the UK MRC, grants MC_UU_00004/07 and MC_UU_00004/09. M.O.H. was supported by the National Heart, Lung, and Blood Institute of the United States National Institutes of Health (NIH) under award R00HL141678. M.O.H. was also supported by the Patient-Centered Outcomes Research Institute (PCORI) Awards (ME-2022C2-27676, ME-2020C1-19220). F.L. was supported by PCORI Awards (ME-2022C2-27676, ME-2020C1-19220). All statements in this report are solely those of the authors and do not necessarily represent the views of the NIH, PCORI, its Board of Governors or Methodology Committee.

**Author contributions**

B.C.K. developed the idea for the manuscript, analysed the data, and wrote the first draft. F.L. assisted with analyses. F.L., B.B, V.J, A.J.C., S.J, and M.O.H. provided feedback on the manuscript structure and content and provided edits. All authors approved of the final submitted manuscript.

**Declaration of Conflicting Interests**

The Authors declare that there is no conflict of interest.

**Data availability**

Data from the RESTORE trial was obtained from the U.S. National Heart, Lung and Blood Institute, Biologic Specimen and Data Repository Information Coordinating Center (BioLINCC) (https://biolincc.nhlbi.nih.gov/home/). The authors do not have the permission to share this data directly; however, the data are available to researchers upon request and approval from BioLINCC.